\documentclass[conference]{IEEEtran}
\IEEEoverridecommandlockouts
\usepackage{cite}
\usepackage{amsmath,amssymb,amsfonts}
\usepackage{algorithmic}
\usepackage{graphicx}
\usepackage{textcomp}
\usepackage{dblfloatfix}
\usepackage{xcolor}
\usepackage{bbding}
\usepackage{makecell}
\usepackage{algorithm}  
\usepackage{color}
\usepackage{multirow}
\usepackage{array}
\usepackage{float}
\usepackage{booktabs}
\usepackage{tabularx,booktabs}
\usepackage{multicol}
\usepackage{cleveref}
\usepackage{balance}
\usepackage{lastpage}
\usepackage{tabulary}
\usepackage{subfigure}
\usepackage{bbm}
\usepackage{mathrsfs}
\usepackage{fancyhdr} 
\usepackage{amsfonts,amssymb}

\def\BibTeX{{\rm B\kern-.05em{\sc i\kern-.025em b}\kern-.08em
    T\kern-.1667em\lower.7ex\hbox{E}\kern-.125emX}}
\def\BibTeX{{\rm B\kern-.05em{\sc i\kern-.025em b}\kern-.08em
    T\kern-.1667em\lower.7ex\hbox{E}\kern-.125emX}}

\begin{document}

\title{Knowledge Transfer based Radio and Computation Resource Allocation for 5G RAN Slicing \\
}
\author{\IEEEauthorblockN{Hao Zhou, and Melike Erol-Kantarci, \IEEEmembership{Senior Member, IEEE}}
\IEEEauthorblockA{\textit{School of Electrical Engineering and Computer Science} \\
\textit{University of Ottawa}\\
Emails:\{hzhou098, melike.erolkantarci\}@uottawa.ca}}

\maketitle

\thispagestyle{fancy} %
      \lhead{} 
      \chead{Accepted by 2022 IEEE Consumer Communications \& Networking Conference, \copyright2022 IEEE 
 } 
      \rhead{} 
      \lfoot{} 
      \cfoot{\thepage} 
      \rfoot{} 
      \renewcommand{\headrulewidth}{0pt} 
      \renewcommand{\footrulewidth}{0pt} 
\pagestyle{fancy}

\begin{abstract}
To implement network slicing in 5G, resource allocation is a key function to allocate limited network resources such as radio and computation resources to multiple slices. However, the joint resource allocation also leads to a higher complexity in the network management. In this work, we propose a knowledge transfer based resource allocation (KTRA) method to jointly allocate radio and computation resources for 5G RAN slicing. Compared with existing works, the main difference is that the proposed KTRA method has a knowledge transfer capability. It is designed to use the prior knowledge of similar tasks to improve performance of the target task, e.g., faster convergence speed or higher average reward. The proposed KTRA is compared with Q-learning based resource allocation (QLRA), and KTRA method presents a 18.4\% lower URLLC delay and a 30.1\% higher eMBB throughput as well as a faster convergence speed.

\end{abstract}

\begin{IEEEkeywords}
Network slicing, 5G, radio access network, mobile edge computation, transfer reinforcement learning.
\end{IEEEkeywords}

\section{Introduction}

Network slicing is an important technique in 5G to enable flexibility and customization. Based on software defined networks and network function virtualization techniques, network slicing can define various virtual network slices over a single physical network infrastructure \cite{b1}. To realize the network slicing, resource allocation is an important part to guarantee service level agreements of slices. The network operator needs to allocate limited resources between slices. For example, enhanced Mobile Broad Band (eMBB) slice usually requires a high throughput, while Ultra Reliable Low Latency Communications (URLLC) slice needs a low latency and a high reliability. Compared with core network slicing, the RAN slicing is more complicated due to limited resources and dynamic channel states \cite{b2}. In \cite{b3}, a risk sensitive model for the resource allocation of URLLC slice is presented. A two-layer method is introduced in \cite{b4} to realize an efficient and low complexity RAN slicing. \cite{b5} shows a RAN slicing scheme by considering both rate and latency demands of various traffic types, and the scheme is tested in a industry 4.0 case. 

The aforementioned works mainly focus on the radio resource allocation. However, the evolving network architecture requires a joint resource allocation scheme, such as radio and computation resources. Indeed, to support the emerging computation intensive applications, deploying mobile edge computing (MEC) servers becomes an ideal solution\cite{b5-1}. The computation tasks can be processed in the MEC server instead of the central cloud, thus a lower delay is achieved. The work of \cite{b6} shows that joint resources allocation slicing has a better performance than slicing one single resource, and \cite{b7} introduced a mathematical model to jointly slice mobile network and edge computation resource.    

Although incorporating computation capability into RAN will bring significant benefits, it also leads to a higher network management complexity, especially when multiple slices are involved. To this end, machine learning methods provide a good opportunity for network management \cite{b8}. For example, in reinforcement learning, the agent interacts with environment to maximize the long term reward based on Markov decision process (MDP), and the complexity of defining a dedicated optimization model is avoided. However, the reinforcement learning algorithms that are applied in most existing works, such as Q-learning and deep Q-learning, generally require a huge number of samples to train the algorithm, which consequently lead to a long convergence time. In addition, the low training efficiency will unavoidably affect the system performance, especially for tasks with tight delay budgets. 
To this end, we propose a knowledge transfer based resource allocation (KTRA) method in this paper. Different with existing algorithms such as reinforcement learning or deep reinforcement learning, the proposed KTRA method has a knowledge transfer capability, and the agent can leverage the knowledge of other expert agents to improve its own performance on the target task \cite{b9}. With the prior knowledge of experts, it requires less samples when exploring the target task, which means a higher exploration efficiency. The proposed KTRA is compared with Q-learning based resource allocation (QLRA), and the simulation shows that KTRA achieves a 18.4\% lower delay for URLLC slice and a 30.1\% higher throughput for eMBB slice as well as a faster convergence.

The rest of this work is organized as follows. Section \ref{s2} presents the related work. Section \ref{s3} introduces the system model and problem formulation, and Section \ref{s4} defines the KTRA scheme and baseline algorithm. We show simulation results in Section \ref{s5}, and conclude this work in Section \ref{s6}.

\section{Related Work}
\label{s2}
Recently machine learning techniques have been extensively studied for wireless network applications, and various techniques are proposed for resource allocation of 5G networks. \cite{b10} proposed a reinforcement learning based method for joint power and radio resource allocation for URLLC and eMBB users. A Q-learning based solution is presented in \cite{b11} to maximize the network utility by satisfying network slicing requests under network resources constraints. \cite{b12} introduced a RAN slicing method to dynamically allocate radio and computation resources, and a constrained learning scheme is defined. A decentralized deep reinforcement learning method is presented in \cite{b13} for network slicing, which ensures service level agreements under networking and computation resources constraints. The joint RAN slicing and computation offloading problem is investigated in \cite{b14}, and multi-agent deep Q-learning is applied to maximize the communication and computation resources utilization. Furthermore, \cite{b14-1} proposed a actor-critic network based deep reinforcement learning method for joint radio and computation resource allocation of virtualized RAN.  

Although various machine learning methods have been proposed for resource allocation of 5G, including reinforcement learning\cite{b10,b11,b12}, deep reinforcement learning\cite{b13,b14-1}, and multi-agent deep reinforcement learning\cite{b14}, these methods usually require a huge amount of samples for training, which means a long exploration phase. Deep reinforcement learning is considered as a breakthrough, but the time-consuming network training is a well known issue \cite{b15}. We propose a correlated Q-learning based method for radio resource allocation of network slicing in \cite{b16}, but the knowledge transfer capability is still not considered. To this end, we propose a KTRA scheme for the joint radio and computation resources allocation of 5G networks in this work. Based on the knowledge transfer strategy, our proposed method can utilize the prior knowledge of expert agent to achieve a faster convergence speed or a higher average reward.

\section{System Model and Problem Formulation}
\label{s3}
\subsection{Network Architecture}

\begin{figure}[!t]
\centering
\includegraphics[width=9cm,height=7cm]{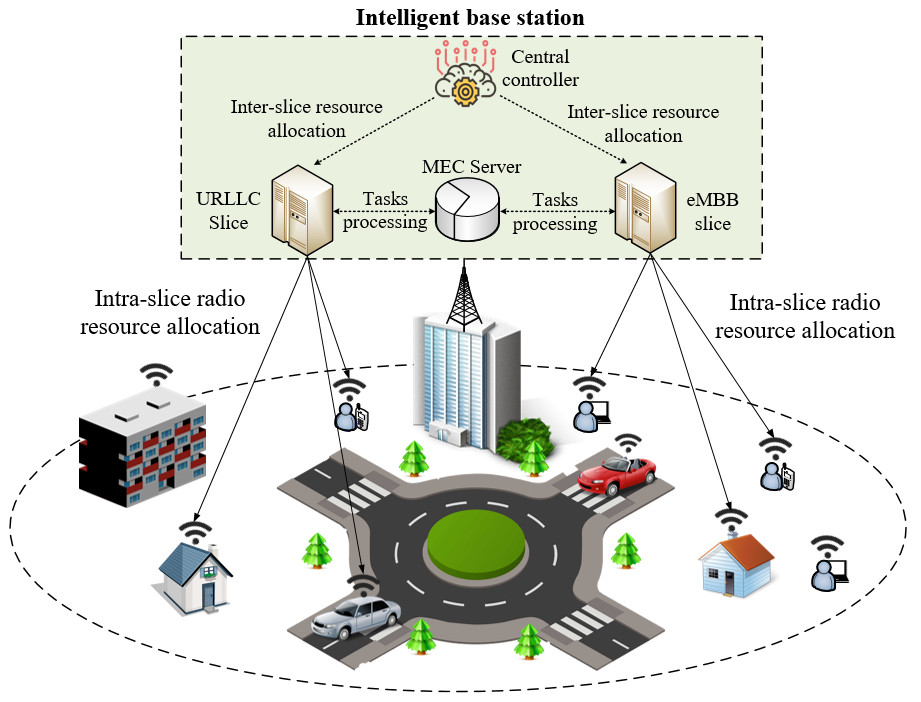}
\caption{Proposed system architecture.}
\label{fig1}
\vspace{-10pt}
\end{figure}

The proposed system model is shown as Fig.\ref{fig1}. We assume the base station (BS) is equipped with a MEC server to process computation tasks of eMBB and URLLC slices. We apply a two-step resource allocation scheme. In the inter-slice phase, the BS intelligently allocates radio and computation resources between different slices. Then these resources are utilized within each slice in the intra-slice phase. For example, the URLLC slice operator will distribute radio resource between attached UEs, and use available MEC server capacity to process the computation tasks of its UEs. In this work, we mainly focus on the inter-slice resource allocation. Given this architecture, the delay experienced by a computation task is:
\begin{equation} \label{eq1}
d=d^{tx}+d^{rtx}+d^{que}+\alpha d^{edge}+(1-\alpha)d^{cloud},
\end{equation}
where $d^{tx}$ and $d^{rtx}$ are the task transmission and retransmission delays, respectively. $d^{que}$ is the queuing delay in BS, which is considered as the scheduling delay. $d^{edge}$ is the processing delay in MEC server, and $d^{cloud}$ is the processing delay in central cloud. $\alpha$ is a binary variable. $\alpha=1$ if the task is computed in the BS, and $\alpha=0$ means the task is forwarded to the central cloud. We presume the intelligent BS will decide whether to process the task in MEC server or offload it to the cloud. Eq. (\ref{eq1}) shows that the delay is affected by both radio and computation resources. The radio resource allocation will affect the transmission delay $d^{tx}$, and computation resource allocation can change the processing delay $d^{edge}$. Meanwhile, the scheduling efficiency will affect the queuing delay $d^{que}$. Following we will explain the communication and computation model.    

\subsection{Communication Model}
We consider resource blocks (RBs) as the smallest time-frequency resource that is distributed to users. The transmission delay $d^{tx}$ between BS and user equipment (UE) is:
\begin{equation} \label{eq2}
d^{tx}=\frac{L_{u}}{E_{j,u}},
\end{equation}
where $L_{u}$ is the transmitted packet size of UE $u$, $E_{j,u}$ is the link capacity between the BS $j$ and the UE $u$. The link capacity depends on the number of RBs that are allocated to this transmission:
\begin{equation}
\resizebox{0.9\hsize}{!}{$\begin{split}
 \label{eq3}
E_{j,u}=&\sum _{r\in{\mathcal{N}_{u}}}b_{RB}log(1+\\ &\frac{p_{j,r}x_{j,u,r}g_{j,u,r}}{b^{RB}N_{0}+\sum\limits_{j'\in \mathcal{J}_{-j}}\sum\limits_{u'\in \mathcal{U}_{j'}}\sum\limits_{r'\in \mathcal{N}_{j'}}{p_{j',r'}x_{j',u',r'}g_{j',u',r'}}}),
\end{split}$}
\end{equation}
where $\mathcal{N}_{u}$ denotes the set of RBs that is allocated to UE $u$, $b_{RB}$ denotes the bandwidth of one RB, $N_{0}$ denotes the noise power density, $p_{j,r}$ denotes the transmission power of RB $r$ in BS $j$. $x_{j,u,r}$ is a binary variable. $x_{j,u,r}=1$ means the RB $r$ is allocated to UE $u$; otherwise $x_{j,u,r}=0$. $g_{j,u,r}$ denotes the channel gain between BS $j$ and UE $u$. $\mathcal{J}_{-j}$ denotes the set of BSs except $j^{th}$ BS, $\mathcal{U}_{j'}$ denotes the set of UEs in BS $j'$, and $\mathcal{N}_{j'}$ denotes the set of total RBs in BS $j'$.   

\subsection{Computation Model}
For a computation task from UEs, we assume it requires certain computation resources to complete the task (denoted by number of CPU cycles). Then the processing delay in MEC server $d^{edge}$ is:
\begin{equation} \label{eq4}
d^{edge}=\frac{c_{u,q}}{\beta C_{j}},
\end{equation}
where $c_{u,q}$ denotes required computation resources of task $q$ from UE $u$, $\beta$ denotes the proportion of computation resources allocated to this task ($0\leq \beta \leq 1$), and $C_{j}$ denotes the total computation capacity of MEC server in BS $j$. 

On the other hand, BS may decide to offload the task to central cloud computation servers. The $d^{cloud}$ in Eq. (\ref{eq1}) is described as:
\begin{equation} \label{eq4-1}
d^{cloud}=d^{up}+d^{down}+d^{c,que}+d^{c,computation},  
\end{equation}
where $d^{up}$ and $d^{down}$ are upload and download transmission delay of the computation task, respectively. $d^{c,que}$ is the cloud queuing delay, and $d^{c,computation}$ is cloud computation delay. The $d^{down}$ and $d^{c,computation}$ can be omitted because: i) the downloaded packet size after computation is usually much smaller than input packet; ii) the central cloud usually has a very high computation capacity \cite{b16-2}. Then 
Eq. (\ref{eq4-1}) can be rewritten as:
\begin{equation} \label{eq5}
d^{cloud}=\frac{1}{\frac{B}{L_{s}}-\lambda}+d^{c, que},
\end{equation}
where $B$ is the backhaul capacity, $B/L_{u}$ denote the service rate, and $\lambda$ is the packet arrival rate. We apply the M/M/1 queue model to describe the upload delay $d^{up}=\frac{1}{\frac{B}{L_{s}}-\lambda}$. Meanwhile, this work mainly focuses the RAN resource allocation, and it is reasonable to assume a fixed cloud queuing delay $d^{c, que}$. Finally, we assume: i) the task is preferred to be processed in the MEC server of the BS due to the potential benefit of MEC; ii) BS will offloaded the task to central cloud if the queuing time expires the preset target delay \cite{b5-1}.         

\subsection{Problem Formulation}

Here we consider two typical slices: eMBB and URLLC slices. The eMBB slice intends to maximize the throughput, while the URLLC slice requires a lower latency. The intelligent BS needs to balance the requirements of both slices, then we define the problem formulation as following: 
\begin{subequations}\label{e2:main}
\begin{align}
\text{max}  \qquad & w^{embb}b^{embb,avg}_{j}+w^{urllc}(d^{tar}-d^{urllc,avg}_{j})& \tag{\ref{e2:main}} \\
 \text{s.t.}  \qquad & b^{embb,avg}_{j}=\frac{\sum\limits_{u\in \mathcal{M}^{embb}_{j}}b^{embb}_{j,u}}{|\mathcal{M}^{embb}_{j}|} & \label{e2:a}  \\
& b^{urllc,avg}_{j}=\frac{\sum\limits_{v\in\mathcal{M}^{urllc}_{j}}d^{urllc}_{j,v}}{|\mathcal{M}^{urllc}_{j}|} & \label{e2:b}  \\
 & (\ref{eq1})\, (\ref{eq2})\, (\ref{eq3}) \, (\ref{eq4})\, (\ref{eq5}) & \label{e2:c}  \\
&\sum\limits_{u\in \mathcal{M}^{embb}_{j}}{x_{j,u,r'}}+\sum\limits_{v\in \mathcal{M}^{urllc}_{j}}{x_{j,v,r'}}=1 & \label{e2:d}\\
  \sum\limits_{r'\in \mathcal{N}_{j'}}  & (\sum\limits_{u\in \mathcal{M}^{embb}_{j}}{x_{j,u,r'}}+\sum\limits_{v\in \mathcal{M}^{urllc}_{j}}{x_{j,v,r'}})\leq |\mathcal{N}_{j}|& \label{e2:e}\\
& C^{embb}_{j}+C^{urllc}_{j}\leq C_{j} & \label{e2:f}
\end{align}
\end{subequations}
where $b^{embb,avg}_{j}$ and $d^{urllc,avg}_{j}$ denote the average throughput and latency of eMBB and URLLC slices, respectively, which are calculated by eq. (\ref{e2:a}) and (\ref{e2:b}). $w^{embb}$ and $w^{urllc}$ are weighting factors to balance two metrics and form a overall objective.  $d^{tar}$ is the preset target delay of URLLC slice, which will bring a positive reward if $d^{tar}>d^{urllc,avg}_{j}$.
$b^{embb}_{j,u}$ is the throughput of UE $u$ in the eMBB slice of BS $j$, and $d^{urllc}_{j,v}$ is the latency of UE $v$ in the URLLC slice. $\mathcal{M}^{embb}_{j}$ and $\mathcal{M}^{urllc}_{j}$ denote the UE set of eMBB and URLLC slices in BS $j$, respectively.   In eq. (\ref{e2:main}), a higher eMBB throughput and a lower URLLC latency are expected to maximize the objective. The eq. (\ref{e2:d}) guarantees one RB can only be allocated to one UE. Eq. (\ref{e2:e}) and (\ref{e2:f}) mean that total allocated RBs and computation capacity should not exceed the available resources in the BS $j$.             

\section{Knowledge Transfer based Radio and Computation Resource Allocation }
\label{s4}
In this section, we first introduce the proposed KTRA method, including the knowledge transfer strategy, MDP definition, and the knowledge transfer reinforcement learning. Then we introduce the Q-learning based resource allocation scheme as a baseline algorithm.  

\subsection{Knowledge Transfer Strategy and MDP definition.}

\begin{figure}[t]
\centering
\includegraphics[width=9cm,height=5cm]{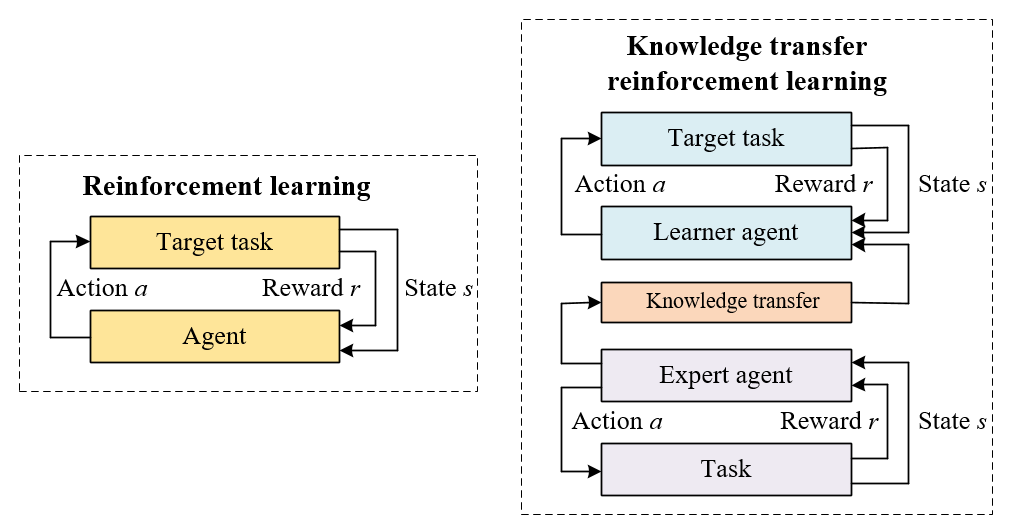}
\caption{Learning strategy comparison .}
\label{fig2}
\vspace{-10pt}
\end{figure}

First, we compare the knowledge transfer reinforcement learning with reinforcement learning to better explain the knowledge transfer strategy. The interaction between agent and task can be described by MDP $<S,A,T,R>$, where $S$, $A$, $T$, $R$ are the set of states, set of actions, transition probability, and reward function, respectively. As shown in Fig.\ref{fig2}, in reinforcement learning, the agent selects an action, receives reward and arrives new state. The agent starts from scratch to explore the task, since no prior knowledge is available in the learning phase.

By contrast, in knowledge transfer reinforcement learning, two agents are involved, namely learner and expert agents. Compared with single learning phase in RL, knowledge transfer reinforcement learning has two phases: knowledge transfer phase and learning phase. In knowledge transfer phase, considering different MDP definitions of two agents, a map function is needed to transform the expert agent's knowledge to the leaner agent. In the learning phase, learner agent utilizes the knowledge of expert agent to improve its own performance on target task. Learner agent is expected to achieve a higher exploration efficiency, since it already has some prior knowledge at the beginning of exploration. 

In this work, we define a learner agent for joint radio and computation resources allocation, and an expert agent for radio resource allocation. The expert agent has no knowledge for computation resource allocation, but it's knowledge of radio resource allocation can be used by learner agent. Following we will define the state, action and reward of two agents. 

\begin{itemize}
  \item \textbf{State}:  The states of both expert and learner agents are $(q^{embb},q^{urllc})$. $q^{embb}$ denotes the number of tasks in the queue of eMBB slice, and $q^{urllc}$ is defined similarly. $(q^{embb},q^{urllc})$ represents the demands of two slices, and agent can select actions accordingly.  
  \item \textbf{Action}: For the expert agent, it only implements the radio resource allocation, and then the action is defined as $(r^{embb},r^{urllc})$, which represents the number of radio resources allocated to eMBB and URLLC slices. For the learner agent, it considers both radio and computation resource allocation, and then the action is defined as $(r^{embb},r^{urllc},c^{embb},c^{urllc})$, where $c^{embb}$ and $c^{urllc}$ denote the computation capacity allocated to eMBB and URLLC slices, respectively.  
  \item \textbf{Reward}: The reward is defined by the objective of slices, which is denoted by (\ref{e2:main}). We also apply a penalty to guarantee the packet drop rate.  
\end{itemize}

\subsection{Knowledge Transfer Reinforcement Learning}
Here we will introduce the knowledge transfer reinforcement learning and define the map function for knowledge transfer. In reinforcement learning, to maximize the expected reward, the agent needs to improve its policy $\pi(s_{t})$ to select the best action at state $s_{t}$, and arrives a new state $s_{t+1}$. Then the state value is defined to represent the potential reward of arriving a new state:
\begin{equation} \label{eu_eqn}
V_{\pi}(s_{t+1}) =\mathbb{E}_{\pi}(\sum_{n=0}^{\infty}\gamma^{n} r_{t+1}|s=s_{t+1}),
\end{equation}
where $V_{\pi}(s_{t+1})$ is the state value to describe the expected reward if the agent arrives $s_{t+1}$, $r_{t+1}$ is the reward at time $t+1$, and $\gamma$ is the discount factor $(0<\gamma<1)$. Furthermore, we need to define the state-action value to describe the expected reward of taking action $a_{t}$ under state $s_{t}$. In Q-learning, the Q-values are updated by:
\begin{equation} \label{eq7}
\begin{aligned}
Q^{new}(s_{t},a_{t}) &= Q^{old}(s_{t},a_{t})+\\
&\alpha(r+\gamma \max\limits_{a} Q(s_{t+1},a)-Q^{old}(s_{t},a_{t})),
\end{aligned}
\end{equation}
where $Q^{old}$ and $Q^{old}$ denote old and new Q-values, respectively, $a_{t}$ is the action at time $t$, $r$ is the reward, and $\alpha$ is the learning rate ($0< \alpha <1$). 

\begin{algorithm}[!t]
	\caption{KTRA algorithm}
	\begin{algorithmic}[1]
		\STATE \textbf{Initialize:} Wireless network parameters and Q-table of the expert.
		\FOR{$TTI=1$ to $T$}
		\FOR{Every BS}
		\STATE With probability $\epsilon$ choose actions randomly, otherwise select $a_{l,t}$ by greedy policy.
		\STATE BS allocates radio and computation resource between slices. Slices process computation tasks by allocated MEC server capacity, and allocate RBs by proportional fairness algorithm.     
		\STATE Calculating reward based on received metrics. 
		\STATE Updating state $s_{l,t}$, find $Q^{T}(\mathcal{F}(s_{l,t}),\mathcal{F'}(a_{l,t}))$.
		\STATE Updating Q-values by eq.(\ref{eq8}).
		\ENDFOR
		\ENDFOR
	\end{algorithmic}
\end{algorithm}

\begin{algorithm}[!t]
	\caption{QLRA algorithm}
	\begin{algorithmic}[1]
		\STATE \textbf{Initialize:} Wireless network parameters. 
		\FOR{$TTI=1$ to $T$}
		\FOR{Every BS}
		\STATE With probability $\epsilon$ choose actions randomly, otherwise select $a^{l,t}$ by greedy policy.
	    \STATE BS allocates radio and computation resource between slices. Slices process computation tasks by allocated MEC server capacity, and allocate RBs by proportional fairness algorithm.
		\STATE Calculating reward based on received metrics. 
		\STATE Updating state $s_{t}$.
		\STATE Updating Q-values by eq.(\ref{eq7}).
		\ENDFOR
		\ENDFOR
	\end{algorithmic}
\end{algorithm}

In Q-learning, the agent needs to explore the state-action space to find the optimal action sequence. This exploration usually takes a large number of iterations, because the agent needs to try huge amounts of action combinations without any prior knowledge. On the contrary, in knowledge transfer reinforcement learning, we use the prior knowledge of experts to improve the exploration efficiency\cite{b17}. The Q-values are updated by:
\begin{equation} \label{eq8}
\resizebox{0.89\hsize}{!}{$\begin{aligned}
Q^{new}(s_{l,t},a_{l,t})=  &Q^{T}(\mathcal{F}(s_{l,t}),\mathcal{F'}(a_{l,t}))+Q^{old}(s_{l,t},a_{l,t})+\\
&\alpha(r+\gamma \max\limits_{a} Q(s_{l,t+1},a)-Q^{old}(s_{l,t},a_{l,t})),
\end{aligned}$}
\end{equation}
where $s_{l,t}$ and $a_{l,t}$ are learner's state and action at time $t$, respectively. $Q^{T}(\mathcal{F}(s_{l,t}),\mathcal{F'}(a_{l,t}))$ is the mapped Q-values as an extra reward of selecting $a_{l,t}$ under state $s_{l,t}$, which aims to guide the exploration of learner agent. $\mathcal{F}$ and $\mathcal{F'}$ are state and action map functions, respectively. $Q^{T}(\mathcal{F}(s_{l,t}),\mathcal{F'}(a_{l,t}))$ can be generated by:
\begin{equation} \label{eq9}
\begin{aligned}
Q^{T}(\mathcal{F}(s_{l,t}),\mathcal{F'}(a_{l,t}))= Q^{e}(s_{e},a_{e}),
\end{aligned}
\end{equation}
where $Q^{e}$ is the Q-values of expert agent, $s_{e}$ and $a_{e}$ are state and action of expert agent. The goal of eq. (\ref{eq9}) is to find a specific Q-value of expert agent to represent the potential reward of taking $a_{l,t}$ under $s_{l,t}$ in learner. Thus we need to find similar states and actions in expert agent's Q-table. Note that the expert and learner agents have the same state definition, then we can always find a $s_{e}$ that satisfy $s_{l,t}=s_{e}$, then $\mathcal{F}$ can be easily defined. Meanwhile, based on the definition of actions, for any $a_{l}=(r^{eM},r^{UR},c^{eM},c^{UR})$, we consider $a_{e}=(r^{eM},r^{UR})$ as a similar action for mapping, and $\mathcal{F'}$ can be defined accordingly.

The KTRA algorithm is summarized as Algorithm 1. Noting that we apply a classic proportional fairness algorithm for intra-slice radio resource allocation, because this work mainly focus on inter-slice level scheduling \cite{b16}, and there is no intra-slice computation resource allocation.

\subsection{Baseline Algorithms: QLRA}
We apply QLRA as a baseline algorithm. Q-learning is the most generally applied reinforcement learning algorithm. The MDP definition of QLRA is the same with KTRA, but there is no prior knowledge. QLRA is given in Algorithm 2. 

\section{Performance Evaluation}
\label{s5}
\subsection{Parameter Settings}
We consider three different cases, including:

\begin{itemize}
  \item \textbf{Case I}:  Q-learning based radio resource allocation. It works as an expert for the Case II.   
  \item \textbf{Case II}: KTRA based joint radio and computation resource allocation. As a learner, BSs in Case II can utilize the Q-tables of expert agent as prior knowledge.  
  \item \textbf{Case III}: QLRA based joint radio and computation resource allocation. It is considered as a baseline algorithm without any prior knowledge for the task.  
\end{itemize}

Each case contains 5 BSs with 500 m inter-site distance, and each BS is considered as an independent agent to implement the proposed strategy. For example, all BSs in Case II will implement the KTRA independently to achieve its own goal. We assume each BS has one eMBB slice with 5 UEs and one URLLC slice with 10UEs. The avaliable bandwidth of one BS is 20 MHz, which contains 100 RBs. We assume there are 13 resource block groups (RBGs) to reduce the allocation complexity. The first 12 RBGs contains 8 RBs each, while the last RBG has 4 RBs. 200 CPU cycles are required to process 1 bit data  \cite{b12}. Other parameters are shown as Table \ref{tab2}.

\begin{table}[!t]
\vspace{-5pt}
\caption{Parameters Settings}
\centering
\renewcommand\arraystretch{1.4}
\begin{tabular}{|p{4cm}<{\centering}||p{3.8cm}<{\centering}|}
\hline
 \textbf{5G Networking} & \textbf{Computation Settings}\\
\hline
3GPP Urban Macro network & Computation capacity: 3 GHz\\ 
 2 OFDM symbols for each TTI & CPU cycles required per bit: 200\\
\cline{2-2}
  Tx/Rx antenna gain: 15 dB.  & \textbf{Traffic Model} \\
\cline{2-2}
  \quad \, Number of subcarriers \quad \quad \, in each RB: 12 & URLLC$\backslash$eMBB traffic: Poisson distribution\\
  Subcarrier bandwidth: 15kHz & URLLC packet size: 50 Bytes \\ 
  Transmission power: 40 dBm & eMBB packet size: 100 Bytes \\ 
  \cline{2-2}
 Backhaul capacity: 10 Mpbs& \textbf{Problem Formulation}\\
\cline{1-2}
 \textbf{Propagation Model} & eMBB$\backslash$URLLC weight factor: 1 \\
 \cline{1-1}
  128.1+37.6log(distance(km)) & URLLC target delay: 2 ms \\
Log-Normal shadowing: 8 dB.& Fixed cloud queuing delay: 1 ms \\
 \cline{1-2}
 \textbf{Retransmission Settings} & \textbf{Learning Settings}\\
\cline{1-2}
Max number of retransmissions: 1  & Learning rate: 0.9 \\
Round trip delay: 4 TTIs &  Discount factor: 0.5 \\
Protocol: asynchronous HARQ. & Epsilon value: 0.05 \\
\hline
\end{tabular}
\label{tab2}
\vspace{-10pt}
\end{table}

\begin{figure*}[t!]
\centering
\label{f3}
\vspace{0pt}
\subfigure[ URLLC latency distribution \text{[ms]} under 2 Mbps eMBB and URLLC traffic, and 3 GHz MEC server per cell.]{
\includegraphics[width=7.4cm,height=5.8cm]{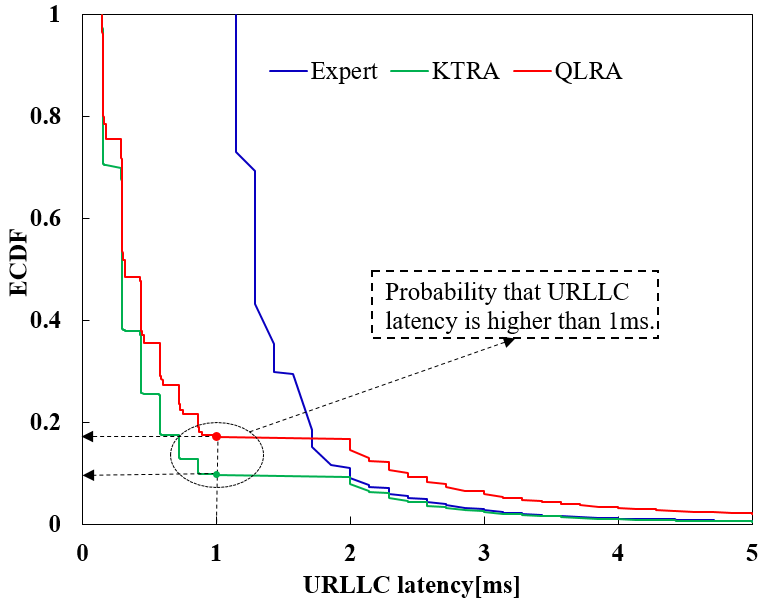}
}
\quad
\subfigure[ Average URLLC latency \text{[ms]} under various URLLC traffic, 2 Mbps eMBB traffic and 3 GHz MEC server per cell.]{
\includegraphics[width=7.4cm,height=5.8cm]{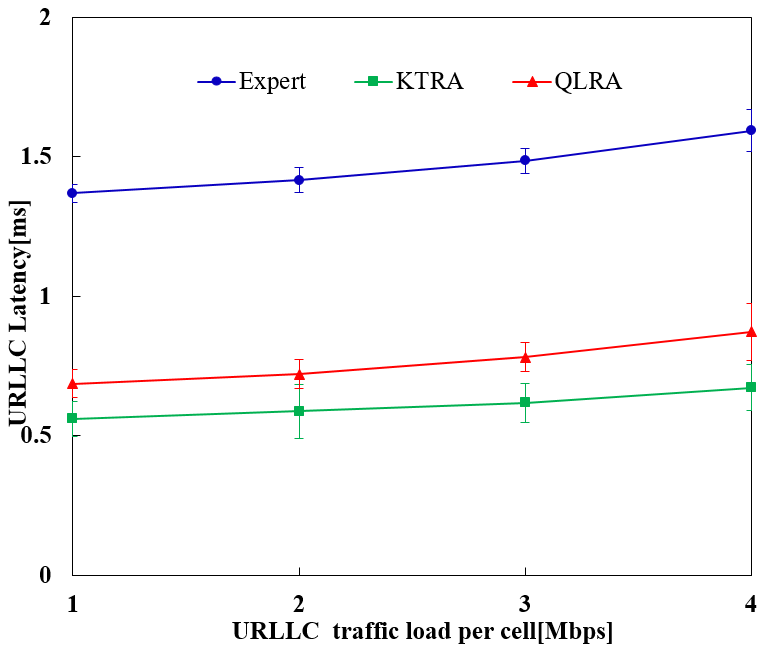}
}
\quad
\subfigure[eMBB throughput \text{[Mbps]} per cell under various URLLC traffic, 2 Mbps eMBB traffic and 3 GHz MEC server.]{
\includegraphics[width=7.4cm,height=5.8cm]{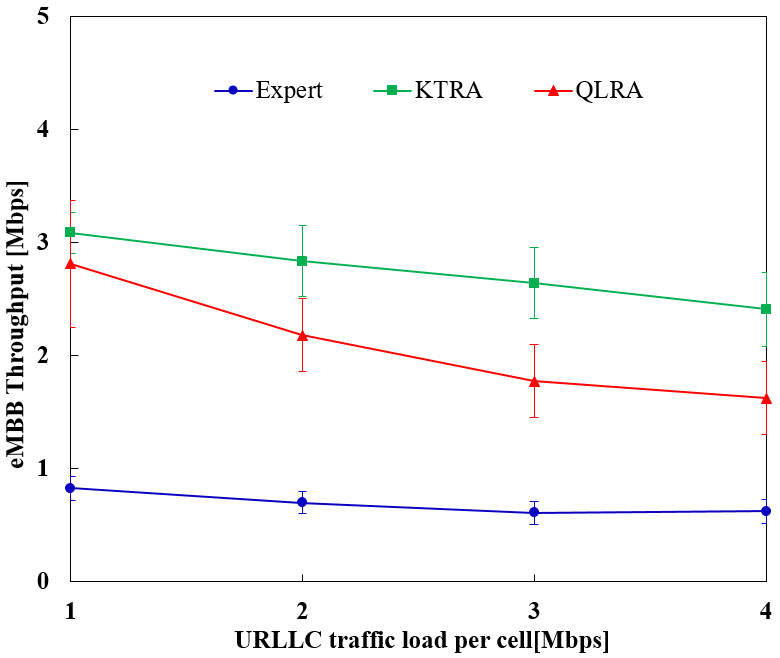}
}
\quad
\subfigure[Average URLLC latency \text{[ms]} under various MEC server capacities, 2 Mbps eMBB and URLLC traffic per cell.]{
\includegraphics[width=7.4cm,height=5.8cm]{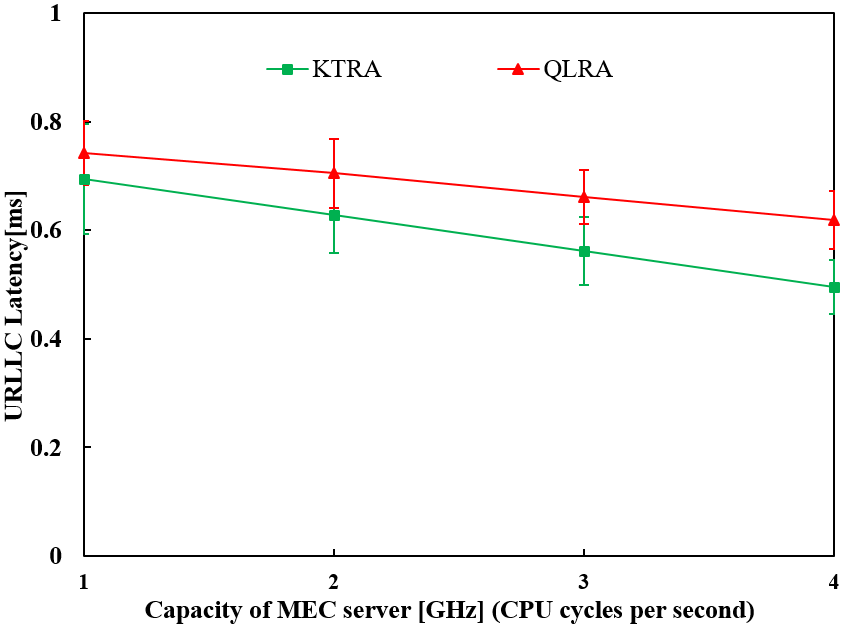}
}
\quad
\subfigure[eMBB throughput \text{[Mbps]} per cell under various MEC server capacities, 2 Mbps eMBB and URLLC traffic.]{
\includegraphics[width=7.4cm,height=5.8cm]{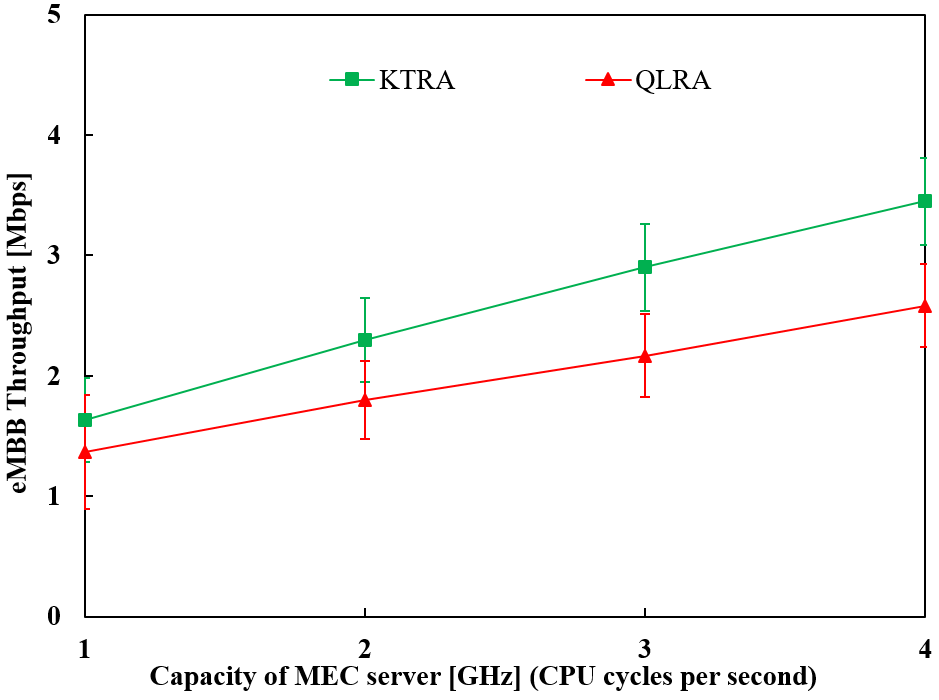}
}
\quad
\subfigure[Convergence performance of KTRA and QLRA under 2 Mbps eMBB and URLLC traffic, and 2 GHz computation capacity per cell.]{
\includegraphics[width=7.4cm,height=5.8cm]{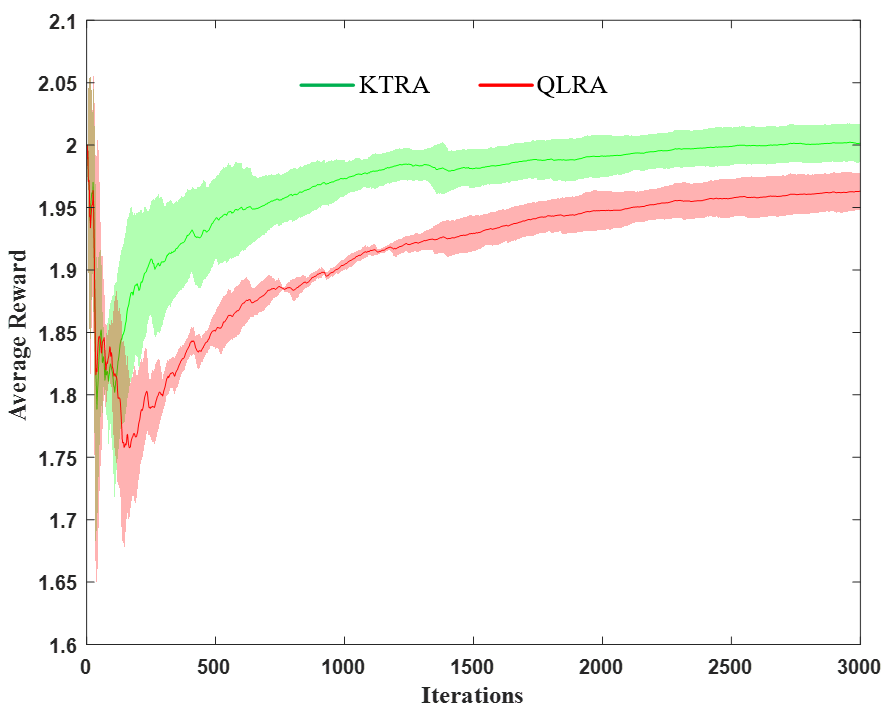}
}
\caption{Simulation results comparison.}
\label{f3}
\end{figure*}

\subsection{Simulation Results}

Firstly, Fig.\ref{f3} (a) shows the empirical cumulative distribution function (ECDF) of URLLC latency of three cases under 2 Mbps eMBB and URLLC traffic, and 3 GHz MEC server per cell. The result shows that expert case has the highest delay distribution, and the reason is that we assume MEC servers are not deployed in expert. Thus all computation tasks need to be processed in the central cloud, which leads to a higher delay. Meanwhile, the proposed KTRA method outperforms QLRA by a better delay distribution, as indicated by a lower ECDF curve in Fig.\ref{f3} (a), and it can be explained by the knowledge transfer capability of KTRA. 

Fig.\ref{f3} (b) presents the average delay experienced by URLLC UEs, and Fig.\ref{f3} (c) shows the average eMBB throughput per cell. Here the eMBB traffic load is fixed to 2 Mbps per cell, and URLLC traffic varies from 1 Mbps to 4 Mbps. Without MEC servers, expert case still has the highest delay and the lowest throughput. Meanwhile, KTRA method achieves a lower URLLC delay and higher eMBB throughput than QLRA. KTRA has a 18.4\% lower URLLC delay and 30.1\% higher eMBB throughput under 2 Mbps URLLC traffic. 

Shown by Fig.\ref{f3} (d) and (e), we investigate the network performance under various MEC server capacities, which is indicated by CPU cycles per second. Considering expert case has no MEC capability, we focus on the performance of KTRA and QLRA to further present the advantage of proposed method. As expected, both algorithms have a lower URLLC delay and a higher eMBB throughput by increasing the MEC server capacity, because higher computation capacity means lower task processing delay. KTRA still outperforms QLRA by a 15.1\% lower URLLC delay and a 33.8\% higher eMBB throughput under 3 GHz MEC server capacity. 

Furthermore, we compare the convergence performance in Fig.\ref{f3} (f). Based on prior knowledge of expert, KTRA has a significantly higher exploration efficiency, which is indicated by a shorter exploration period and a higher average reward. On the contrary, QLRA suffers a longer exploration phase and a lower average reward, because it needs to explore the task from scratch. To summarize, KTRA achieves a better performance in both network metrics (higher URLLC delay and lower eMBB throughput) and machine learning metrics (better convergence speed and higher average reward). 

Finally, the packet drop rate of KTRA and QLRA are 0.042\% and 0.048\%, respectively, under 2 Mpbs eMBB traffic and 4 Mbps URLLC traffic. The satisfying packet drop rate is because we apply a penalty in both algorithms to prevent dropping packet.     

\section{Conclusion}
\label{s6}
The evolving network architecture requires more efficient solutions for network resource allocation. In this work, we propose a KTRA method for joint radio and computation resource allocation. Compared with existing works, the main difference is that the proposed method has a knowledge transfer capability. The proposed KTRA method is compared with Q-learning based resource allocation, and KTRA presents a 18.4\% lower URLLC delay and 30.1\% higher eMBB throughput as well as a faster convergence. In the future, we will consider the knowledge transfer of tasks with different state definition.

\section*{Acknowledgment}
This work is supported by Natural Sciences and Engineering Research Council of Canada (NSERC), Collaborative Research and Training Experience Program (CREATE) under Grant 497981 and Canada Research Chairs Program.


\begin{thebibliography}{00}

\bibitem{b1} A. Ksentini, and N. Nikaein, ”Toward Enforcing Network Slicing on RAN: Flexibility and Resources Abstraction,” \textit{IEEE Communications Magazine}, vol. 55, no. 6, pp.102-108, Jun. 2017.

\bibitem{b2} A. A. Barakabitze, A. Ahmad, R. Mijumbi, and A. Hinesd, “5G network slicing using SDN and NFV: A survey of taxonomy, architectures and future challenges,”  \textit{Computer Networks}, vol. 167, pp.1-40, Feb. 2020.

\bibitem{b3} M. Alsenwi, N. Tran, M. Bennis, A. Bairagi, and C. Hong, “eMBBURLLC
Resource Slicing: A Risk-Sensitive Approach,” \textit{IEEE Communications
Letters}, vol. 23, no. 4, pp. 740-743, Apr. 2019.

\bibitem{b4} D. Marabissi, and R. Fantacci, ”Highly Flexible RAN Slicing Approach
to Manage Isolation, Priority, Efficiency,” \textit{IEEE Access}, vol.7, pp.97130-97142, Jul. 2019.

\bibitem{b5} J. G. Morales, M. C. L. Estan, and J. Gozalvez, "Latency-Sensitive 5G RAN Slicing for Industry 4.0," \textit{IEEE Access}, vol.7, pp. 143139-143159, Sep.2019. 

\bibitem{b5-1} N. Abbas, Y. Zhang, A. Taherkordi, and T. Skeie, "Mobile Edge Computing: A Survey," \textit{IEEE Internet of Things Journal}, vol. 5, no. 1, pp. 450-465 Feb. 2018. 

\bibitem{b6} A. Huang, Y. Li, Y. Xiao, X. Ge, S. Sun, and H. Chao, "Distributed Resource Allocation for Network Slicing of Bandwidth and Computational Resource," \textit{in Proceedings of 2020 IEEE International Conference on Communications}, pp.1-6, Jun. 2020.

\bibitem{b7} B. Xiang, J. Elias, F. Martignon, and E. D. Nitto, "Joint network slicing and mobile edge computing in 5G networks," \textit{in Proceedings of 2019 IEEE International Conference on Communications}, pp.1-6, Jul. 2019.

\bibitem{b8} M. Elsayed and M. Erol-Kantarci, “AI-Enabled Future Wireless Networks:
Challenges, Opportunities, and Open Issues,” \textit{IEEE Vehicular Technology Magazine}, vol. 14, no.3, pp. 70-77, Sep.2019. 

\bibitem{b9} M. E. Taylor, P. Stone, and Y. Liu, ”Transfer Learning for Reinforcement Learning Domains: A Survey,” \textit{Journal of Machine Learning Research}, vol.10, pp. 1633-1685, Sep. 2009.

\bibitem{b10} M. Elsayed and M. Erol-Kantarci, “AI-Enabled Radio Resource Allocation
in 5G for URLLC and eMBB Users,” \textit{in Proceedings of 2019 IEEE 2nd 5G World Forum (5GWF)}, pp. 590-595, Sep. 2019. 

\bibitem{b11} Y. Shi, Y. E. Sagduyu, and T. Erpek, "Reinforcement Learning for Dynamic Resource Optimization in 5G Radio Access Network Slicing," \textit{in Proceedings of 2020 IEEE 25th International Workshop on Computer Aided Modeling and Design of Communication Links and Networks}, pp. 1-6, Sep. 2020. 

\bibitem{b12} W. Wu, N. Chen, C. Zhou, M. Li, X. Shen, W. Zhuang and X. Li, "Dynamic RAN Slicing for Service-Oriented Vehicular Networks via Constrained Learning," \textit{IEEE Journal on Selected Areas in Communications}, vol. 39, no. 7, pp. 2076-2089, Jul. 2021. 

\bibitem{b13} Q. Liu, T. Han, and E. Moges, "EdgeSlice: Slicing Wireless Edge Computing Network with Decentralized Deep Reinforcement Learning," \textit{in Proceedings of 2020 IEEE International Conference on Distributed Computing Systems}, pp.1-11, Dec. 2020.

\bibitem{b14} Q. Ye, W. Shi, K. Qu, H. He, W. Zhuang, and X. Shen, "Joint RAN Slicing and Computation Offloading for Autonomous Vehicular Networks: A Learning-Assisted Hierarchical Approach," \textit{IEEE Open Journal of Vehicular Technology}, vol.2, pp. 272-288, Jun. 2021.

\bibitem{b14-1} J. A. Ayala-Romero, A. Garcia-Saavedra, M. Gramaglia, X. Costa-Perez, A. Banchs and J. J. Alcaraz, "vrAIn: Deep Learning based Orchestration for Computing and Radio Resources in vRANs," \textit{IEEE Transactions on Mobile Computing (early access)}, doi: 10.1109/TMC.2020.3043100, Dec. 2020. 

\bibitem{b15} V. Mnih et al., “Human-Level Control Through Deep Reinforcement Learning,” \textit{Nature}, pp. 529–533, Feb. 2015.

\bibitem{b16} H. Zhou, M. Elsayed, and M. Erol-Kantarci, “RAN Resource Slicing in 5G Using Multi-Agent Correlated Q-Learning,” arXiv:2107.01018 [cs.NI], Jun. 2021. 

\bibitem{b16-2} Y. K. Tun, D. H. Kim, M. Alsenwi, N. H. Tran, Z. Han, and C. S. Hong, "Energy Efficient Communication and Computation Resource Slicing for eMBB and URLLC Coexistence in 5G and Beyond," \textit{IEEE Access}, vol.8, pp. 136024-136035, Jul. 2020. 

\bibitem{b17} M. Elsayed, M. Erol-Kantarci, and H. Yanikomeroglu, ”Transfer Reinforcement Learning for 5G New Radio mmWave Networks,” \textit{IEEE Transaction on Wireless Communications}, vol.20, no.5, pp.2838-2849, May. 2021. 

\end{thebibliography}
\end{document}